\documentclass[twoside,twocolumn,9pt]{article}
\usepackage{extsizes}
\usepackage[super,sort&compress,comma]{natbib} 
\usepackage[version=3]{mhchem}
\usepackage[left=1.5cm, right=1.5cm, top=1.785cm, bottom=2.0cm]{geometry}
\usepackage{physics, siunitx}
\usepackage{amsmath, amssymb, amsthm} 
\usepackage{balance}
\usepackage{mathptmx}
\usepackage{sectsty}
\usepackage{graphicx} 
\usepackage{lastpage}
\usepackage[format=plain,justification=justified,singlelinecheck=false,font={stretch=1.125,small,sf},labelfont=bf,labelsep=space]{caption}
\usepackage{float, subfig}
\usepackage{fancyhdr}
\usepackage{fnpos}
\usepackage[english]{babel}
\addto{\captionsenglish}{%
  
}
\usepackage{lmodern}
\usepackage{array}
\usepackage{droidsans}
\usepackage{charter}
\usepackage[T1]{fontenc}
\usepackage[usenames,dvipsnames]{xcolor}
\usepackage{setspace}
\usepackage[compact]{titlesec}
\usepackage{hyperref}
\hypersetup{
    pdfauthor={Amartya Bose},
    pdftitle={Effect of Temperature Gradient on Quantum Transport},
    colorlinks=true,
    allcolors=.
}

\usepackage{epstopdf}

\definecolor{cream}{RGB}{222,217,201}

\begin{document}

\pagestyle{fancy}
\thispagestyle{plain}
\fancypagestyle{plain}{
    \renewcommand{\headrulewidth}{0pt}
}

\makeFNbottom
\makeatletter
\renewcommand\LARGE{\@setfontsize\LARGE{15pt}{17}}
\renewcommand\Large{\@setfontsize\Large{12pt}{14}}
\renewcommand\large{\@setfontsize\large{10pt}{12}}
\renewcommand\footnotesize{\@setfontsize\footnotesize{7pt}{10}}
\renewcommand\scriptsize{\@setfontsize\scriptsize{7pt}{7}}
\makeatother

\renewcommand{\thefootnote}{\fnsymbol{footnote}}
\renewcommand\footnoterule{\vspace*{1pt}%
    \color{cream}\hrule width 3.5in height 0.4pt \color{black} \vspace*{5pt}}
\setcounter{secnumdepth}{5}

\makeatletter
\renewcommand\@biblabel[1]{#1}
\renewcommand\@makefntext[1]%
{\noindent\makebox[0pt][r]{\@thefnmark\,}#1}
\makeatother
\renewcommand{\figurename}{\small{Fig.}~}
\sectionfont{\sffamily\Large}
\subsectionfont{\normalsize}
\subsubsectionfont{\bf}
\setstretch{1.125} 
\setlength{\skip\footins}{0.8cm}
\setlength{\footnotesep}{0.25cm}
\setlength{\jot}{10pt}
\titlespacing*{\section}{0pt}{4pt}{4pt}
\titlespacing*{\subsection}{0pt}{15pt}{1pt}

\fancyfoot{}
\fancyfoot[LO,RE]{\vspace{-7.1pt}\includegraphics[height=9pt]{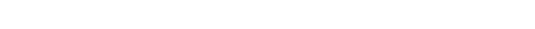}}
\fancyfoot[CO]{\vspace{-7.1pt}\hspace{13.2cm}\includegraphics{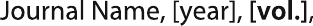}}
\fancyfoot[CE]{\vspace{-7.2pt}\hspace{-14.2cm}\includegraphics{head_foot/RF}}
\fancyfoot[RO]{\footnotesize{\sffamily{1--\pageref{LastPage} ~\textbar  \hspace{2pt}\thepage}}}
\fancyfoot[LE]{\footnotesize{\sffamily{\thepage~\textbar\hspace{3.45cm} 1--\pageref{LastPage}}}}
\fancyhead{}
\renewcommand{\headrulewidth}{0pt}
\renewcommand{\footrulewidth}{0pt}
\setlength{\arrayrulewidth}{1pt}
\setlength{\columnsep}{6.5mm}
\setlength\bibsep{1pt}

\makeatletter
\newlength{\figrulesep}
\setlength{\figrulesep}{0.5\textfloatsep}

\newcommand{\topfigrule}{\vspace*{-1pt}%
    \noindent{\color{cream}\rule[-\figrulesep]{\columnwidth}{1.5pt}} }

\newcommand{\botfigrule}{\vspace*{-2pt}%
    \noindent{\color{cream}\rule[\figrulesep]{\columnwidth}{1.5pt}} }

\newcommand{\dblfigrule}{\vspace*{-1pt}%
    \noindent{\color{cream}\rule[-\figrulesep]{\textwidth}{1.5pt}} }

\makeatother

\twocolumn[
    \begin{@twocolumnfalse}
        {\includegraphics[height=30pt]{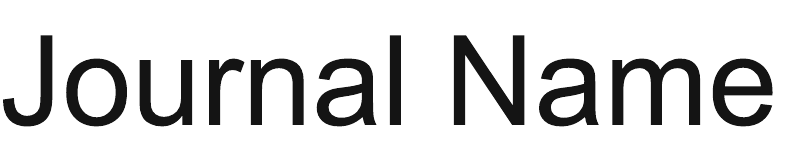}\hfill\raisebox{0pt}[0pt][0pt]{\includegraphics[height=55pt]{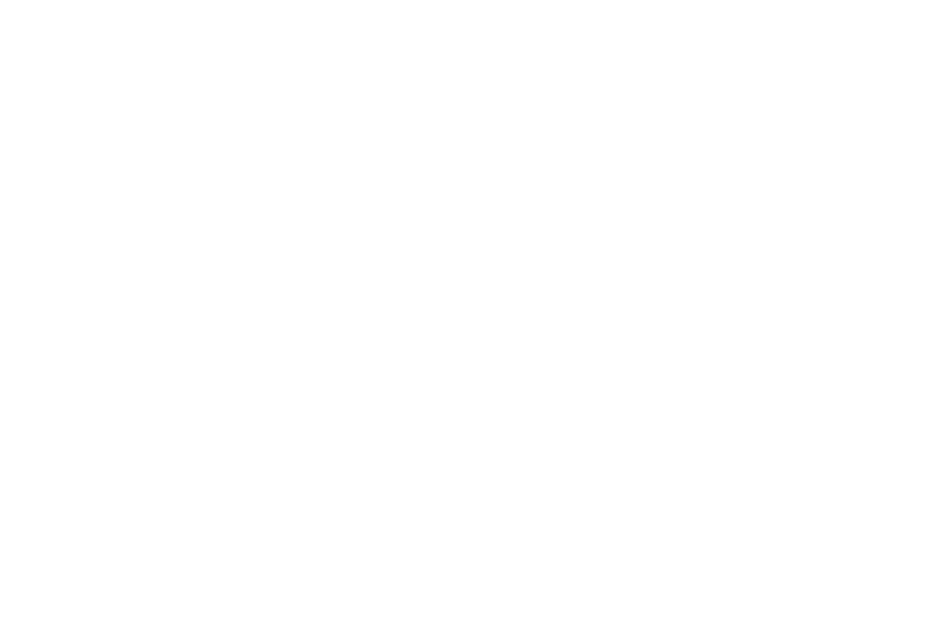}}\\[1ex]
            \includegraphics[width=18.5cm]{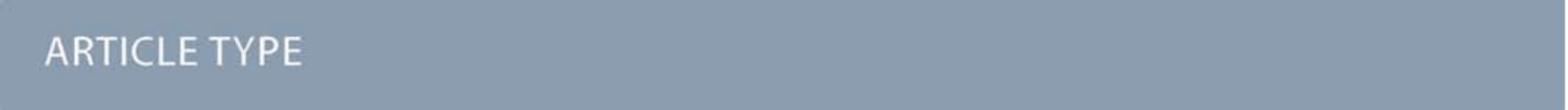}}\par
        \vspace{1em}
        \sffamily
        \begin{tabular}{m{4.5cm} p{13.5cm} }

            \includegraphics{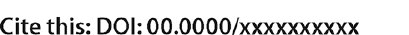}   & \noindent\LARGE{\textbf{Effect of Temperature Gradient on Quantum Transport}}                \\
                                              & \vspace{0.3cm}                                                                               \\

                                              & \noindent\large{Amartya Bose,$^{\ast}$\textit{$^{a}$} and Peter L. Walters\textit{$^{b,c}$}} \\

            \includegraphics{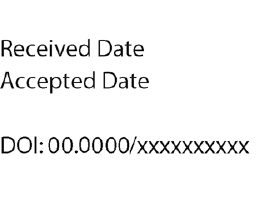} &                                                                                              \\
        \end{tabular}

    \end{@twocolumnfalse} \vspace{0.6cm}

]

\renewcommand*\rmdefault{bch}\normalfont\upshape
\rmfamily
\section*{}
\vspace{-1cm}


\footnotetext{\textit{$^{a}$~Department of Chemistry, Princeton University, Princeton, New Jersey 08544; E-mail: amartyab@princeton.edu, amartya.bose@gmail.com}}
\footnotetext{\textit{$^{b}$~Department of Chemistry, University of California, Berkeley, California 94720.}}
\footnotetext{\textit{$^{c}$~Miller Institute for Basic Research in Science, University of California Berkeley, Berkeley, California 94720.; E-mail: peter.l.walters2@gmail.com}}




\sffamily{\textbf{The recently introduced multisite tensor network path integral (MS-TNPI) method [Bose and Walters, \textit{J. Chem. Phys.}, 2022, \textbf{156}, 24101.] for simulation of quantum dynamics of extended systems has been shown to be effective in studying one-dimensional systems. Quantum transport in these systems are typically studied at a constant temperature. However, temperature seems to be a very obvious parameter that can be spatially changed to control the quantum transport. Here, MS-TNPI is used to study ``non-equilibrium'' effects of an externally imposed temperature gradient on the quantum transport in one-dimensional extended quantum systems.}}


\rmfamily 


Quantum transport in extended open systems has been one of the holy-grails of
quantum dynamics. It combines the difficulty of treating extended quantum
systems with the difficulty of treating open quantum systems, both of which
potentially lead to exponential growth of computational complexity. However,
such systems are ubiquitous in nature, and hence of great importance. From
magnetic materials to molecular aggregates, a vast variety of interesting
physical phenomena lend themselves to be modeled as extended one-dimensional
quantum systems interacting with an open thermal environment. Wave
function-based methods such as density matrix renormalization
group~\cite{whiteDensityMatrixFormulation1992, whiteRealTimeEvolutionUsing2004,
    schollwockDensitymatrixRenormalizationGroup2005,
    schollwockDensitymatrixRenormalizationGroup2011,
    renTimeDependentDensityMatrix2018,
    paeckelTimeevolutionMethodsMatrixproduct2019} (DMRG) and multi-configuration
time-dependent Hartree~\cite{beckMulticonfigurationTimedependentHartree2000,
    wangMultilayerFormulationMulticonfiguration2003,
    maioloMultilayerGaussianBased2021} (MCTDH and ML-MCTDH) and related methods
have proven to be exceptionally useful in simulating the dynamics of extended
quantum systems. However, due to their computational complexity, these methods
are typically less useful when it comes to simulations pertaining to open
quantum systems.

Path integrals have often been presented as a viable solution to the problem of
calculating and storing of the wave functions for these open quantum systems.
With these methods, the main challenge is that the number of paths considered
in the path integral increases exponentially with the number of time steps.
However, this exponential proliferation of the system path list can be
curtailed through the use of an iterative procedure that exploits the rapid
decay of correlation between well-separated time points. Although the
computational complexity still increases exponentially with the number of time
points retained within memory ($L$), this is usually much smaller then the
number of points in the simulation. The quasi-adiabatic propagator path
integral~\cite{makriTensorPropagatorIterative1995,
    makriTensorPropagatorIterative1995a,
    makriModularPathIntegral2018,makriSmallMatrixPath2020} (QuAPI) methods, which
are based on Feynman-Vernon influence
functional,~\cite{feynmanTheoryGeneralQuantum1963} make simulations of general
open quantum systems much more approachable. Of late, the usage of tensor
networks to facilitate simulations with influence functionals has also become
quite common.\cite{strathearnEfficientNonMarkovianQuantum2018,
    jorgensenExploitingCausalTensor2019, boseTensorNetworkRepresentation2021,
    bosePairwiseConnectedTensor2022} Ideas from these tensor network-based influence
functional methods have motivated a recent extension of DMRG to simulating the
dynamics of extended open quantum
systems.\cite{boseMultisiteDecompositionTensor2022} This multisite tensor
network path integral (MS-TNPI) method has also been used to explore the
dynamics and spectrum of the B850 ring of the light harvesting
subsystem\cite{boseTensorNetworkPath2022} and study the effects of phononic scattering on quantum transport.\cite{boseEffectPhononsImpurities2022}

Path integrals for extended open quantum systems suffer from a huge problem of
exponential scaling. First, the presence of the thermal environment leads to an
exponential scaling with respect to the number of time steps within memory. The
base of this exponential scaling is related to the dimensionality of the system.
For extended systems, the dimensionality is the product of the dimensions of the
individual ``sites'' or ``entities.'' So, for a system with say 16 two-level
systems, like the one used to model the B850 ring in
Ref.~\citep{boseTensorNetworkPath2022}, the base of the scaling is
$2^{16}$. So, if the memory length is $L$, the computational complexity is
effectively $2^{16\times 2\times L}$. This is the problem that MS-TNPI solves by
using a DMRG-like decomposition of the system along the various sites along with
a decomposition of the ``paths'' along the temporal dimension.

\begin{figure}
    \centering
    \includegraphics[scale=0.25]{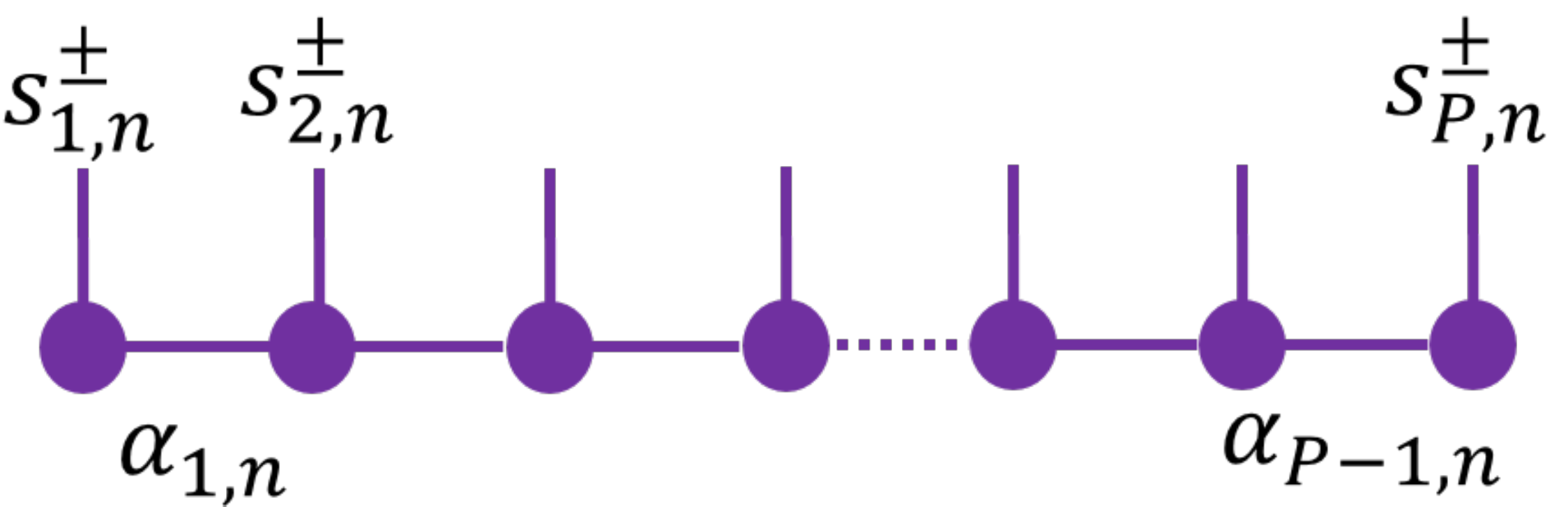}
    \caption{Schematic of a density matrix of an extended system represented as an MPS.}\label{fig:rhoMPS}
\end{figure}

DMRG and DMRG-like methods proceed by decomposing the system along the spatial
dimension. By exploiting the lack of correlation between distant sites, the
resulting matrix product state (MPS) can be an extremely compact and efficient
representation of the system. The reduced density matrix after the
$n$\textsuperscript{th} time step can expressed in the form of an MPS as follows
\begin{align}
    \tilde\rho(S^\pm_n, n\Delta t) & = \sum_{\left\{\alpha_{(j,n)}\right\}} A_{\alpha_{(1,n)}}^{s^\pm_{1,n}} A_{\alpha_{(1,n)},\alpha_{(2,n)}}^{s^\pm_{2,n}}\cdots A_{\alpha_{(P-1,n)}}^{s^\pm_{P,n}},\label{eq:rhoMPS}
\end{align}
where $\alpha_{j,n}$ is the index connecting the $j$\textsuperscript{th} site at
time-step $n$ to the $(j+1)$\textsuperscript{th} site at the same time step.
Figure~\ref{fig:rhoMPS} gives a graphical representation of this structure. In
the notation used here, the forward-backward state of the
$j$\textsuperscript{th} site at the $n$\textsuperscript{th} time point is
denoted by $s^\pm_{j,n}$ and the states of all the sites at this time step are
collectively represented by $S_n^\pm$.  (Here, the forward-backward state is a
combination of the forward, bra, and backward, ket, states of the density
matrix.) When the density matrix is represented as an MPS, the forward-backward
propagator, which evolves it in time, must be represented as a matrix product
operator (MPO).

With this setup, it is possible to obtain the time-dynamics of the isolated
system through a series of MPO-MPS applications.  However, often the individual
sites interact with separate dissipative environments:
\begin{align}
    \hat{H} & = \hat{H}_0 + \sum_{j=1}^{P} \sum_{l=1}^{N_\text{osc}} \frac{p_{jl}^2}{2m_{jl}} + \frac{1}{2}m_{jl}\omega_{jl}^2\left(x_{jl} - \frac{c_{jl} \hat{s}_j}{m_{jl}\omega_{jl}^2}\right)^2
\end{align}
where $\hat{H}_0$ is the Hamiltonian corresponding to the isolated extended
system with $P$ units or particles. Each unit or particle is coupled to a
dissipative environment, which under Gaussian response theory can be mapped on
to a set of $N_\text{osc}$ harmonic oscillators. The $l$\textsuperscript{th}
harmonic oscillator of the $j$\textsuperscript{th} system unit interacts with
it through the operator $\hat{s}_j$ with a strength of $c_{jl}$. The
frequencies and couplings of the baths are often given in terms of the spectral
density, which can be related to the Fourier transform of the energy gap
autocorrelation function.

In the the presence of the dissipative environment, the time evolution of the
reduced density matrix can be described as
\begin{align}
    \tilde\rho(S^\pm_N, N\Delta t) & = \sum_{S_0^\pm}\sum_{S^\pm_1}\cdots\sum_{S^\pm_{N-1}} \tilde\rho(S^\pm_0, 0) P_{S_0^\pm\cdots S^\pm_N}                                            \\ \label{eq:RDM_PathAmp}
                                   & =\sum_{S_0^\pm}\sum_{S^\pm_1}\cdots\sum_{S^\pm_{N-1}} \tilde\rho(S^\pm_0, 0) P^{(0)}_{S_0^\pm\cdots S^\pm_N} F\left[\left\{S^\pm_n\right\}\right].
\end{align}
Here, $P^{(0)}_{S_0^\pm\cdots S^\pm_N}$ is the bare path amplitude tensor, which
contains the full information of the isolated system, and $F$ is the
Feynman-Vernon influence functional~\cite{feynmanTheoryGeneralQuantum1963},
which depends entirely on the spectral density and encodes the
system-environment interaction. Lastly, $P_{S_0^\pm\cdots S^\pm_N}$ is the path
amplitude tensor, which describes the system in the presence of the solvent.
Since the dimensionality of the path amplitude tensor grows exponentially with
the number of particles and time steps, it can only be explicitly constructed in
a very small number of cases.

\begin{figure}
    \centering
    \includegraphics[scale=0.25]{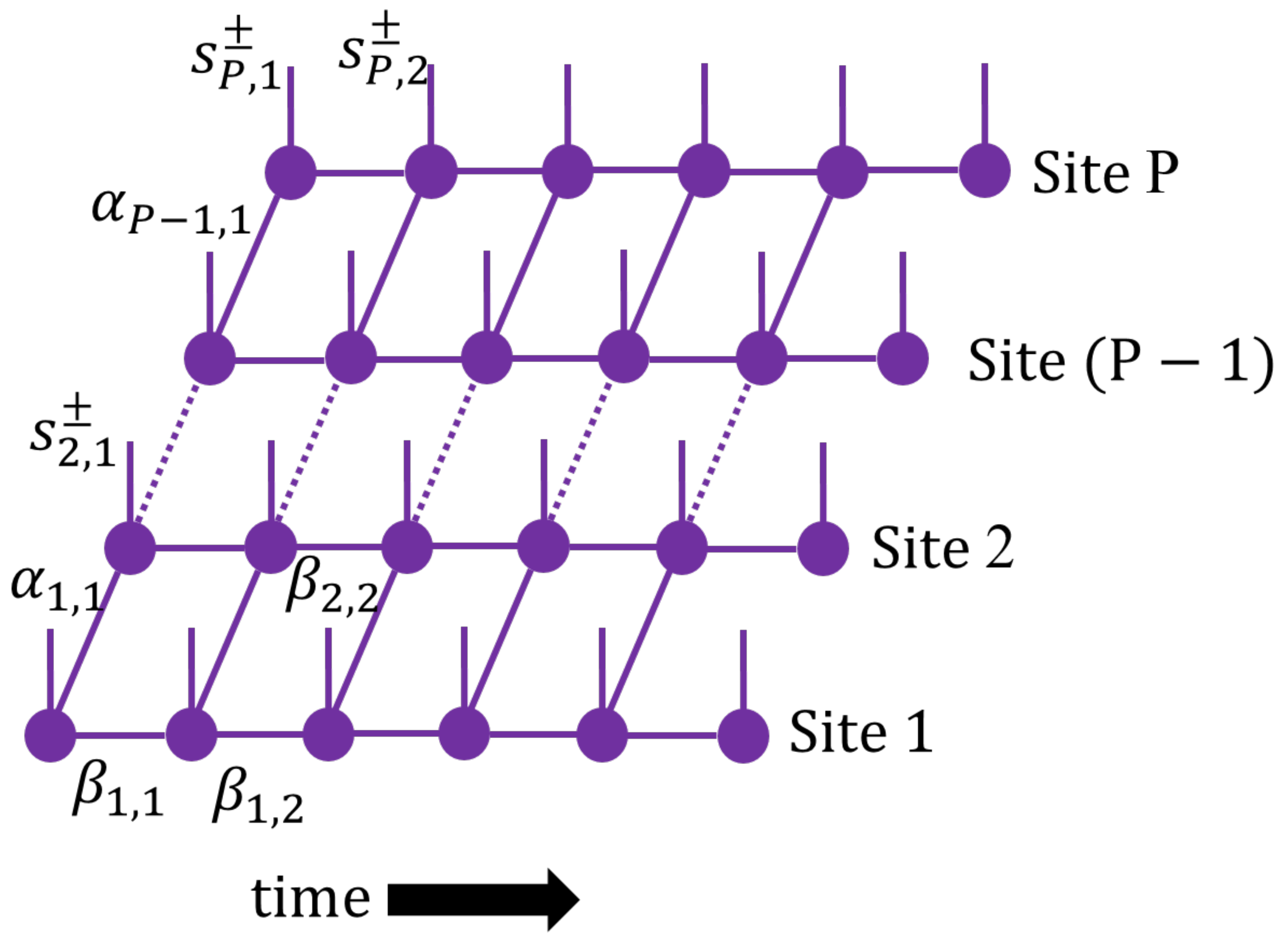}
    \caption{Schematic of the two-dimensional MS-TNPI tensor network.}\label{fig:ms-tnpi}
\end{figure}

\begin{figure}
    \centering
    \includegraphics[scale=0.25]{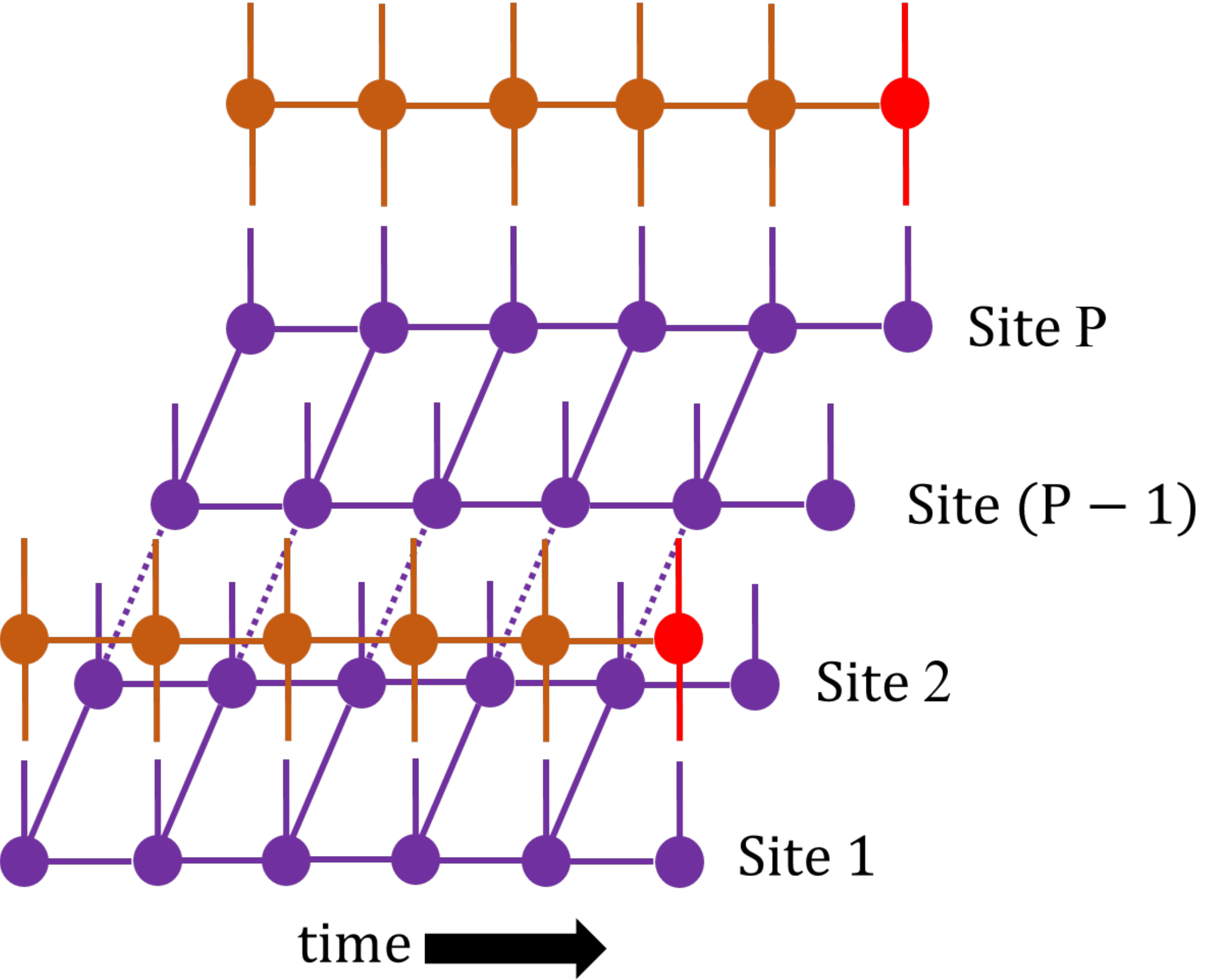}
    \caption{Schematic of incorporation of influence functional in the MS-TNPI tensor network shown only for the first and last sites.}\label{fig:ms-tnpi-if}
\end{figure}

MS-TNPI~\cite{boseMultisiteDecompositionTensor2022} avoids this exponential
scaling by performing a spatial decomposition of the bare system and combining
it with a temporal decomposition of the influence functional to produce a
compact two-dimensional tensor network representation of the path amplitude
tensor,
\begin{align}
    P_{S_{0}^{\pm}\cdots S_{N}^{\pm}} & = \sum_{\left\{\beta_{n}\right\}} \mathbb{T}^{S_{0}^{\pm}}_{\beta_{0}}
    \cdots \mathbb{T}^{S_{n}^{\pm}}_{\beta_{n-1}, \beta_{n}}
    \cdots\mathbb{T}_{\beta_{N-1}}^{S_{N}^{\pm}}.\label{eq:MS-PA_MPS}
\end{align}
Here, $\beta_n$ is the index connecting the tensors at time-point $n$ to the
ones at $n+1$, and each $\mathbb{T}$ is a matrix product representation
decomposed along the site axis. The resulting two-dimensional tensor network is
shown in Fig.~\ref{fig:ms-tnpi}.  Each of the columns roughly contains the state
of the full system at any point of time. Therefore, when contracting the network
along the columns, we get the full reduced density matrix corresponding to the
extended system. Na\"ively speaking, the number of columns in the MS-TNPI
network corresponds to the total length of the simulation. However, an iterative
procedure can be employed that effectively reduces the number columns to the
length of the memory induced by the baths. The rows represent the path amplitude
corresponding to the individual sites or units of the system. This allows both
the incorporation of the Feynman-Vernon influence functional in a transparent
manner as MPOs acting on the rows as shown in Fig.~\ref{fig:ms-tnpi-if} and the
truncation of memory.

The present paper uses MS-TNPI to explore the effects of temperature
differences on quantum transport in extended open systems. Such temperature
differences can be caused by an external temperature gradient being applied
across a molecular wire or more commonly as a side-effect of heat generated
during excitation caused by lasers. Irrespective of the origin of such
temperature differences, it provides us with a potentially useful parameter for
controlling and changing the characteristics of the quantum transport. This is a
first step in exploring such changes brought onto the population dynamics caused
by such temperature gradients. In this work, we focus exclusively on the Frenkel
exciton transfer model. The system Hamiltonian is given by
\begin{align}
    \hat{H}_0 & = \epsilon\sum_{j=1}^{P}\dyad{e_j} + \hbar J\sum_{j=1}^{P}\left(\dyad{e_{j}}{e_{j+1}} + \dyad{e_j}{e_{j+1}}\right)
\end{align}
where $J$ is the excitonic coupling between the different sites, $\epsilon$ is
the excitation energy of the sites and lastly, $\ket{e_j}$ is the many-body
wavefunction with just the $j$\textsuperscript{th} site excited. In terms of the
one-body ground, $\ket{\phi^g_j}$, and excited, $\ket{\phi^e_j}$, wavefunctions,
$\ket{e_j} =  \ket{\phi^e_j} \otimes \prod_{k\ne j} \ket{\phi^g_k}$. The
system-bath interaction takes place through $\hat{s}_j$, which defined as
$\hat{s}_j\ket{\phi^e_j} = \ket{\phi^e_j}$ and $\hat{s}_j\ket{\phi^g_j} = 0$.

\begin{figure}
    \centering
    \includegraphics{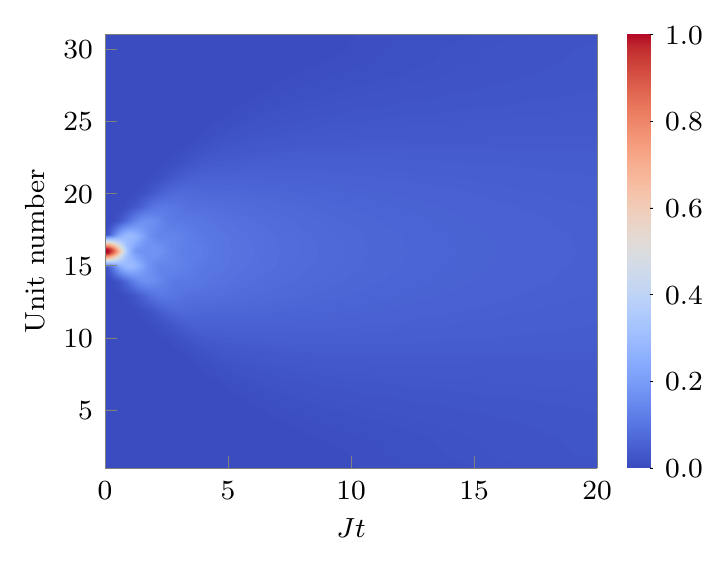}
    \caption{Unit dependent excited state population, $P^\text{exc}_j(t)$, when no temperature gradient is applied.}\label{fig:site_popln}
\end{figure}

As our first example, we consider a 31 site system with $\hbar J=1$ and
$\epsilon =100$. (The sites are numbered from 1 to 31.) The vibrational bath is
site independent and characterized by the following spectral density:
\begin{align}
    J(\omega) & = \frac{\pi}{2}\hbar\xi\omega\exp\left(-\frac{\omega}{\omega_c}\right)
\end{align}
where $\omega_c = 8J$ is the cutoff frequency and the dimensional Kondo
parameter, $\xi = 0.075$. We will study the dynamics of the system under an
average temperature of $k_B \bar{T} = \hbar J$ and a temperature gradient of 0
and $0.05\hbar J / k_B$ per site. The temperature is lowest at the bottom end
where the units have lower numbers and rises as we move up. The system with
a zero temperature gradient is going to be used as a reference for comparisons.
The excited state population dynamics, $P^\text{exc}_j(t) =
    \mel{\phi^e_j}{\tilde\rho(t)}{\phi^e_j}$, corresponding to an initial state of
$\tilde\rho(0) = \dyad{e_{16}}$, in the absence of any temperature gradient is
presented in Fig.~\ref{fig:site_popln}. Because the middle unit is initially
excited, the dynamics is completely symmetric, that is the populations of the
units equidistant from the edges are identical. To explore this symmetry of
dynamics, a site-symmetrized excited state population and a deviation measure
are defined as follows:
\begin{align}
    \bar{P}^\text{exc}_{j}(t) & = \frac{P^\text{exc}_{j}(t) + P^\text{exc}_{P-j+1}(t)}{2}                                 \\
    \delta{P}^\text{exc}_j(t) & = \frac{P^\text{exc}_j(t) - \bar{P}^\text{exc}_j(t)}{\bar{P}^\text{exc}_j(t)} \times 100.
\end{align}

\begin{figure}
    \centering
    \subfloat[$\delta{P}^\text{exc}_j(t)$ in absence of any temperature gradient.]{\includegraphics{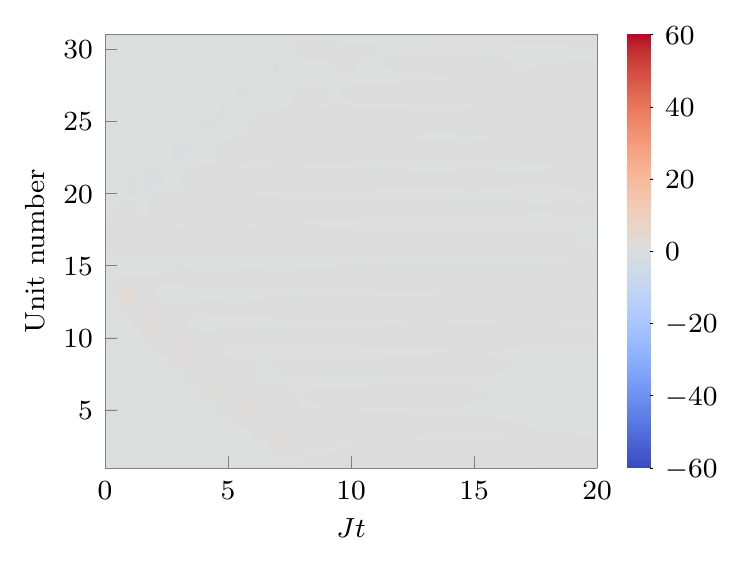}}

    \subfloat[$\delta{P}^\text{exc}_j(t)$ in presence of a temperature gradient of $0.05\hbar J/k_B$ per site.]{\includegraphics{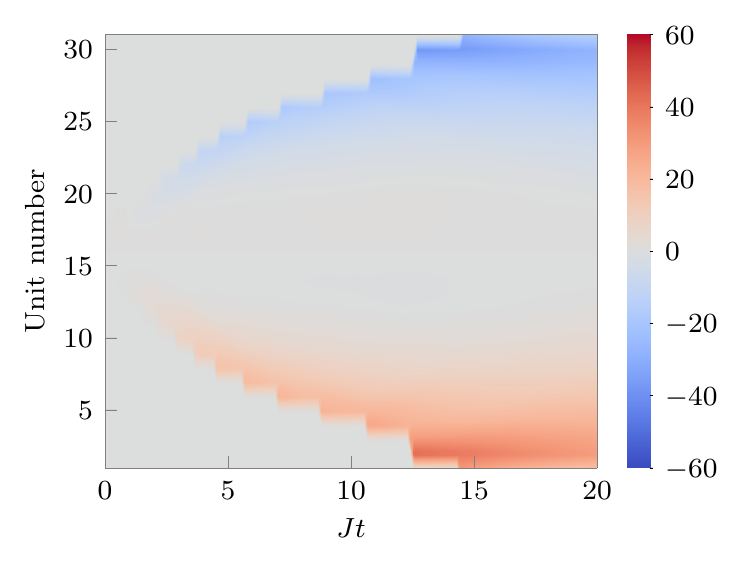}}
    \caption{Plots of $\delta{P}^\text{exc}_j(t)$ in presence and absence of a temperature gradient for Frenkel model coupled to Ohmic bath.}\label{fig:example1_avgdiff}
\end{figure}

This is demonstrated in Fig.~\ref{fig:example1_avgdiff}~(a). The imposition of
an external temperature gradient breaks this symmetry, leading to deviations in
the excited state population dynamics as shown in
Fig.~\ref{fig:example1_avgdiff}~(b). For the linear ramp considered, the
transport process seems to preferentially move the exciton to the colder
monomers. The deviations are quite significant with an upper limit of around
$\pm 50\%$.

\begin{figure}
    \centering
    \includegraphics{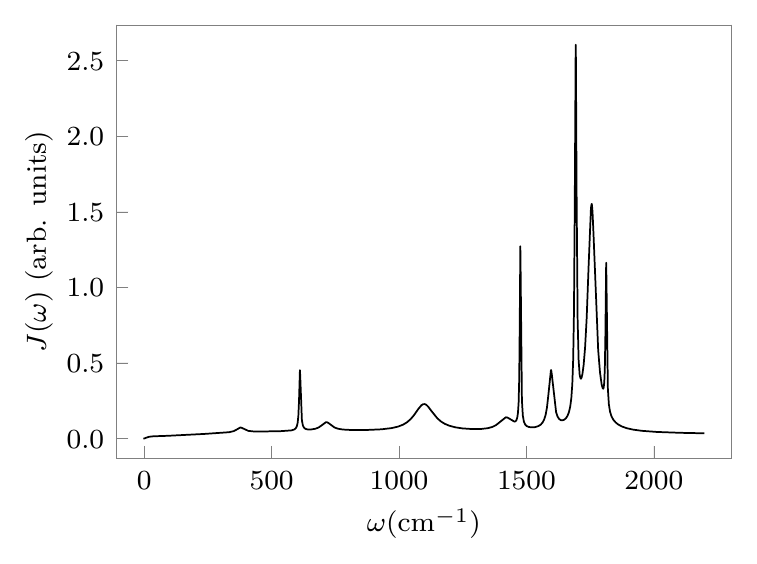}
    \caption{Spectral density describing the molecular vibrations and the impact of the proteins on the dynamics of the B850 ring from Ref.~\cite{olbrichTimeDependentAtomisticView2010}}\label{fig:b850_jw}
\end{figure}

As a more realistic example of exciton transfer, consider a chain of 31
bacteriochlorophyll (BChl) units. The intermonomer electronic coupling is taken
to be $\SI{156.5}{\per\cm}$ and the excitation energy of a BChl unit is taken
to be $\SI{12390}{\per\cm}$. The local spectral density was calculated using the
molecular dynamics-based (MD) bath response function, $C_\text{MD}(t)$,
reported in Ref.~\cite{olbrichTimeDependentAtomisticView2010} using the
following relation:
\begin{align}
    J(\omega) & = \frac{\hbar\omega\beta_\text{MD}}{2}\int_0^\infty C_\text{MD}(t) \cos(\omega t) \dd{t}.\label{eq:harmon_spect}
\end{align}
The resultant spectral density is shown in Fig.~\ref{fig:b850_jw}. Note that the
inverse temperature used here, $\beta_\text{MD}$, corresponds to MD simulation
setup. The spectral density, Eq.~\ref{eq:harmon_spect}, is independent of the
temperature of the path integral simulations done here. So, just as in the
previous example, all the monomer units have identical spectral densities here
as well. The dynamics and absorption spectrum of the B850 ring with this
MD-based solvent has already been studied using
MS-TNPI.~\cite{boseTensorNetworkPath2022} We want to understand the changes
brought about in the dynamics through an external temperature gradient of
$\SI{10}{\kelvin\per\text{unit}}$. The average temperature of the chain is held
at $\SI{300}{\kelvin}$.

\begin{figure}
    \centering
    \includegraphics{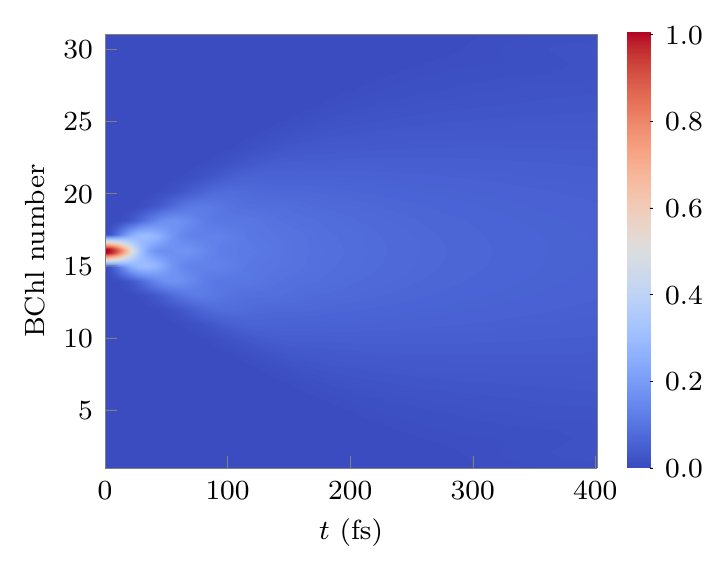}
    \caption{Excitonic population dynamics in a BChl chain in absence of temperature gradient.}\label{fig:bchl_center}
\end{figure}

\begin{figure}
    \centering
    \includegraphics{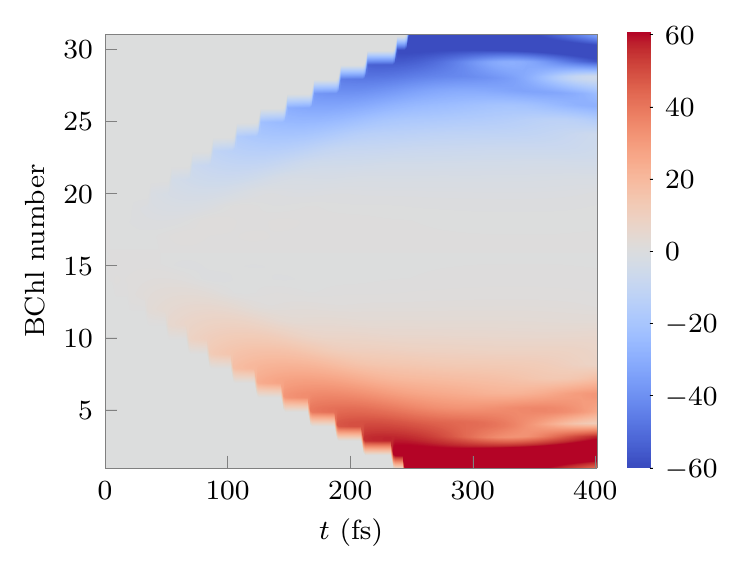}
    \caption{Difference between excitonic population dynamics for a system with temperature gradient and the same system without a temperature gradient.}\label{fig:BChl_popln_diff}
\end{figure}

To study the effect of an external temperature gradient on this system, let us
focus once again on exciting the middle bacteriochlorophyll unit (unit 16). In
absence of any external temperature gradient, the populations of the sites is symmetric at all times. This dynamics is shown in Fig.~\ref{fig:bchl_center}.
Figure~\ref{fig:BChl_popln_diff} shows the difference caused in the excitonic
population dynamics by the externally imposed temperature gradient. Note that
despite having a structured spectral density, the trends here are identical to
the model example, Fig.~\ref{fig:example1_avgdiff}. Even in this case,
population preferentially moves towards the colder end of the chain.

\begin{figure}
    \centering
    \includegraphics{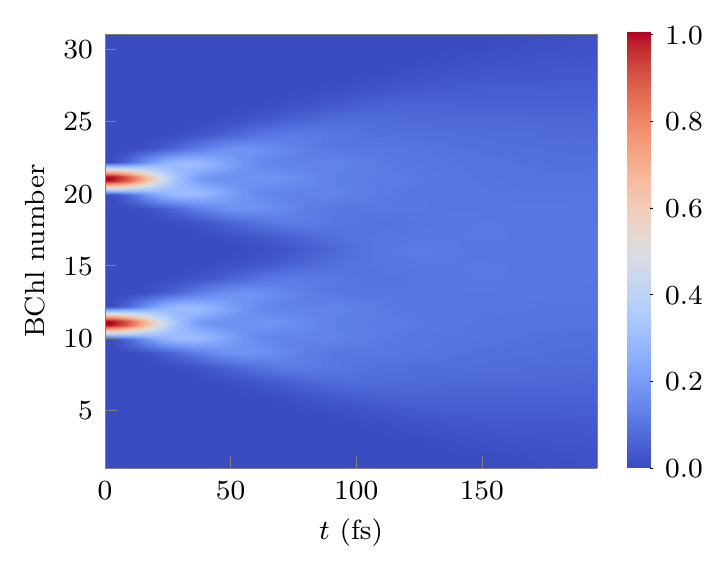}
    \caption{Excitonic population dynamics in a BChl chain starting from a state with two excitons on the 6\textsuperscript{th} and the 10\textsuperscript{th} monomers.}\label{fig:BChl_two_excite_popln}
\end{figure}

\begin{figure}
    \centering
    \includegraphics{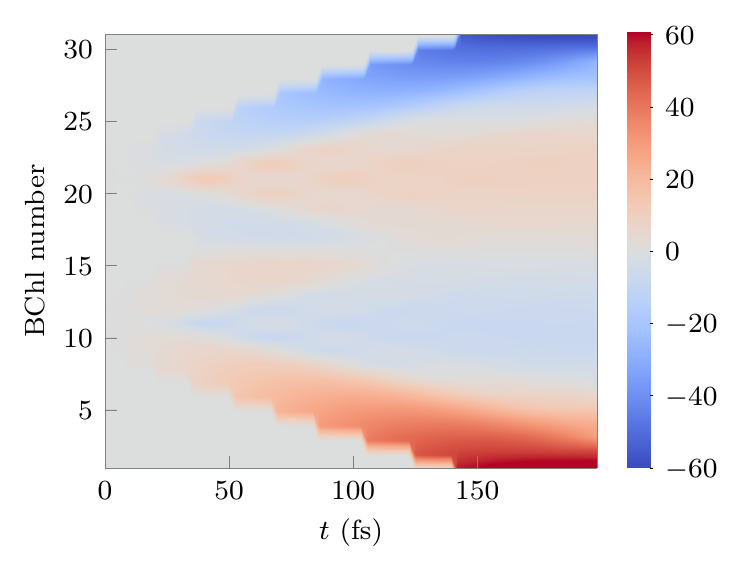}
    \caption{Difference between the excitonic population dynamics in a BChl chain starting from a state with two excitons on the 6\textsuperscript{th} and the 10\textsuperscript{th} monomers in presenece and absence of temperature gradient.}\label{fig:BChl_two_excite_popdiff}
\end{figure}

Consider next the possibility of multiple excitons in the system. Suppose we
start from the initial state $\tilde\rho = \dyad{e_{11}e_{21}}$. So, the
11\textsuperscript{th} and the 21\textsuperscript{th} monomers are in the
excited states and everything else is in the ground state. The dynamics of this
system in absence of any temperature gradient is shown in
Fig.~\ref{fig:BChl_two_excite_popln}. On imposing the same external temperature
gradient, the transport of both the excitons get affected. The excitonic
population difference from the transport in absence of the temperature gradient
is shown in Fig.~\ref{fig:BChl_two_excite_popdiff}. This time, we predictably
see a superposition of the same pattern as the single excitation case for each
of the two centers of excitation. The region between the two initial
excitations, that is monomers 11 and 21 show this interplay.

We have demonstrated a noticeable change in the quantum transport of excitons in
the presence of a temperature gradient. The excitonic population seems to travel
preferentially to the colder end of the chain. This trend is consistent between
model spectral densities and structured ones derived from molecular dynamics
simulations. Investigations on the effect of temperature on the spectra and
other properties will also be undertaken in the future. What has been shown in
this communication seems to indicate that temperature might be a useable control
for quantum dynamics.  Different temperature profiles may affect the dynamics in
interesting ways, opening up possibilities of using it to promote desired
outcomes, like an increased quantum yield on a site, in the dynamics. Further
exploration of these aspects would be done in the future.

\section*{Author Contributions}
Both authors contributed equally to the design, implementation and writing of
this paper.

\section*{Conflicts of interest}
There are no conflicts to declare.

\section*{Acknowledgments}
A.\,B. acknowledges the support of the Computational Chemical Science Center:
Chemistry in Solution and at Interfaces funded by the US Department of Energy
under Award No. DE-SC0019394. P.\,W. acknowledges the Miller Institute for
Basic Research in Science for funding.

\scriptsize{
    \bibliography{library}
    \bibliographystyle{rsc}
}
\end{document}